\documentclass[a4paper,11pt]{article}
\usepackage{jcappub, bm, color} 
\usepackage{amssymb,amsfonts,slashed,amsthm,amsmath,graphicx, soul}
\usepackage[caption=false]{subfig}
\newcommand{\lmk}{\left(}  
\newcommand{\rmk}{\right)}
\newcommand{\lkk}{\left[}  
\newcommand{\rkk}{\right]}

\newcommand{\del}{\partial}

\newcommand{\bea}{\begin{array}}
\newcommand{\eea}{\end{array}}
\newcommand{\beq}{\begin{eqnarray}}
\newcommand{\eeq}{\end{eqnarray}}
\newcommand{\eq}[1]{Eq.~(\ref{#1})}

\newcommand{\abs}[1]{\left\vert {#1} \right\vert}

\newcommand{\M}{M}

\newcommand{\eff}{\Delta N_{\rm eff}}
\newcommand{\neff}{\Delta N_{\rm eff}}
\newcommand{\cc}{\Omega_\Lambda}
\newcommand{\Mpc}{\ {\rm Mpc}}
\newcommand{\Msolar}{M_\odot}

\title{
Anthropic Bound on Dark Radiation 
and its Implications for Reheating
}

\author{
Fuminobu Takahashi$^{\spadesuit \diamondsuit}$
}
\affiliation{$^\spadesuit$ Department of Physics, Tohoku University, 
Sendai, Miyagi 980-8578, Japan} 
\affiliation{$^\diamondsuit$ Kavli IPMU (WPI), UTIAS, 
The University of Tokyo, 
Kashiwa, Chiba 277-8583, Japan}

\author{
and Masaki Yamada$^{\heartsuit}$
}
\affiliation{
$^{\heartsuit}$ Institute of Cosmology, Department of Physics and Astronomy, 
Tufts University, Medford, MA 02155, USA
}

\abstract{
We derive an anthropic bound on the extra neutrino species, $\Delta N_{\rm eff}$, based on the observation that a positive $\Delta N_{\rm eff}$ suppresses the growth of matter fluctuations due to the prolonged radiation dominated era, which reduces the fraction of matter that collapses into galaxies, hence, the number of observers. We vary $\Delta N_{\rm eff}$ and the positive cosmological constant while fixing the other cosmological parameters. We then show that the probability of finding ourselves in a universe satisfying the current bound is of order a few percents for a flat prior distribution. If $\Delta N_{\rm eff}$ is found to be close to the current upper bound or the value suggested by the $H_0$ tension, the anthropic explanation is not very unlikely. On the other hand, if the upper bound on $\Delta N_{\rm eff}$ is significantly improved by future observations, such simple anthropic consideration does not explain the small value of $\Delta N_{\rm eff}$. We also study simple models where dark radiation consists of relativistic particles produced by heavy scalar decays, and show that the prior probability distribution sensitively depends on the number of the particle species. 
}

\begin{document}

\maketitle
\flushbottom

%%%%%%%%%%%%%%%%%%%%%%%%%%%%%%%%%%%%%%%%%%%%%%%%%%%%%%%%%%%%%%%%
\section{Introduction
\label{sec:introduction}}
%%%%%%%%%%%%%%%%%%%%%%%%%%%%%%%%%%%%%%%%%%%%%%%%%%%%%%%%%%%%%%%%

The $\Lambda$CDM paradigm has been hugely successful in explaining various cosmological observations with high accuracy. Remarkably, with only six parameters, it gives a very nice fit to the observed cosmic microwave background (CMB) temperature and polarization anisotropies~\cite{Aghanim:2018eyx}. 

Recently, however,  the $\Lambda$CDM paradigm is challenged by the findings of possible tensions among different observations. In particular, there seems to be a rather clear tension in the estimate of the Hubble constant,  $H_0$, In other words, the Hubble constant measured locally is higher than the value inferred from the Planck CMB observation based on the $\Lambda$CDM model. The recent improved analysis of the local measurements of $H_0$ strengthened the tension to be $4.4 \sigma$~\cite{Riess:2019cxk}.
While it is not trivial to entirely remove the $H_0$ tension by introducing new physics without invoking other tensions, there are several ways that can ameliorate the tension~\cite{Renk:2017rzu,Khosravi:2017hfi,Aylor:2018drw,Dhawan:2017ywl,Mortsell:2018mfj,DEramo:2018vss,Kreisch:2019yzn,Barenboim:2019tux,Pandey:2019plg,Kaloper:2019lpl,Agrawal:2019lmo,Alexander:2019rsc}. One of such extensions is to introduce new relativistic particles, the so-called dark radiation. It is customary to express the amount of dark radiation in terms of the extra neutrino species, $\Delta N_{\rm eff}$. One needs $\Delta N_{\rm eff} \gtrsim 0.4-0.5$ to reduce the $H_0$ tension significantly~\cite{Mortsell:2018mfj,Riess:2019cxk}.

There are a variety of candidates for dark radiation.
In most of the scenarios, dark radiation consists of unknown massless or extremely light particles such as sterile neutrinos, axions,  hidden photons, etc.  The existence (or non-existence) of dark radiation has rich implications for physics beyond the SM as well as the evolution of the early Universe. For instance, if dark radiation was in thermal equilibrium with the
standard model (SM) particles, they must have sizable couplings that can be constrained by direct search experiments or astrophysics~\cite{Nakayama:2010vs,Weinberg:2013kea,Jeong:2013eza, Kawasaki:2015ofa, Chacko:2015noa}. 
On the other hand, dark radiation may be
produced non-thermally by the decay of heavy particles
(see e.g. Refs.~\cite{Ichikawa:2007jv,Cicoli:2012aq,Higaki:2012ar,Higaki:2013lra,Cicoli:2018cgu}).  Indeed, in the string theory, there often appear many light hidden particles (such as axions and hidden photons), and if the inflaton is universally coupled to the light particles including the SM ones, we expect that the Universe is likely filled with hidden particles, which is not consistent with what we observe~\cite{Cicoli:2010ha}. Therefore, if the existence of dark radiation is ubiquitous in the landscape, there may be some reason to suppress its abundance. 

In this Letter, we examine an anthropic explanation of the dark radiation under the assumption that $\Delta N_{\rm eff}$
is an environmental parameter which takes random values  in the multiverse. A similar assumption is made in the anthropic explanation of the observed small cosmological constant~\cite{Davies, Barrow, Barrow:1988yia, Hawking:1987en, Weinberg:1987dv}. 
Specifically, we vary both $\Delta N_{\rm eff}$ (or $N_{\rm eff}$) and the positive cosmological constant while fixing the other cosmological parameters. 
Although we do not give a rigorous
UV completion that provides such a mechanism to distribute different values of $\Delta N_{\rm eff}$, it is possible to imagine that 
the abundances of such light particles depend on their couplings with the inflaton, which may depend on the
choice of the universe. 
We shall study simple toy models along this line, and show that the prior distribution of $\Delta N_{\rm eff}$ sensitively depends on the number of  particle species that constitute dark radiation.
Since it is notoriously difficult to quantify various anthropic conditions, we will adopt a very simple ansatz which seems to be successful in explaining the observed cosmological constant~\cite{Weinberg:1987dv, Martel:1997vi}: the
number of observers in a universe is proportional to the fraction of matter that collapses into galaxies. 
In fact, we note that one can extend the anthropic argument on the cosmological constant to derive
the anthropic bound on $\Delta N_{\rm eff}$ and its likely values. In this sense, our anthropic explanation of dark radiation is on the same footing with that of the cosmological constant.

%%%%%%%%%%%%%%%%%%%%%%%%%%%%%%%%%%%%%%%%%%%%%%%%%%%%%%%%%%%%%%%%
\section{Anthropic bound on dark radiation
\label{sec:anthropic}}
%%%%%%%%%%%%%%%%%%%%%%%%%%%%%%%%%%%%%%%%%%%%%%%%%%%%%%%%%%%%%%%%

\subsection{Probability distribution of $\Delta N_{\rm eff}$ and $\cc$}

The effective neutrino number in the standard cosmology is $N_{\rm eff}^{\rm (std)} \simeq 3.046$.
The energy density of dark radiation $\rho_{\rm DR}$ is conveniently described by a change 
of the effective neutrino number $\Delta N_{\rm eff} \equiv N_{\rm eff} - N_{\rm eff}^{\rm (std)}$ as 
\beq
 \Delta N_{\rm eff} = \frac{4}{7} \frac{\rho_{\rm DR}}{(\pi^2 / 30) T_\nu^4}, 
 \label{delta N_eff}
\eeq
where $T_\nu$ is the neutrino temperature. 
We can express the radiation density parameter, $\Omega_{\rm rad}$, as a function of $\neff$: 
\begin{align}
 \Omega_{\rm rad} \simeq \Omega_{\rm rad}^{\rm (std)} \times 
 \lmk 1 + 0.13 \Delta N_{\rm eff} \rmk, 
 \label{Omega_rad}
\end{align}
where $\Omega_{\rm rad}^{\rm (std)} \simeq 4.18 \times 10^{-5} h^{-2}$ is 
the radiation density parameter in the standard cosmology.

In this Letter, we calculate the conditional probability distribution of $\Delta N_{\rm eff}$ and the density parameter of the cosmological constant $\Omega_\Lambda$ 
in the multiverse, assuming that the probability is proportional to the number of observers in each universe. 
It is estimated by~\cite{Pogosian:2004hd}
\beq
 P(\neff, \cc) \propto P_{\rm prior} (\neff, \cc) 
 \int d M n_{\rm G} (\neff, \cc, M) N_{\rm obs} (\neff, \cc, M), 
 \label{Ptot1}
\eeq
where 
$n_{\rm G}\, dM$ is the comoving number density of galaxies with mass between $M$ and $M + dM$, 
and $N_{\rm obs}$ is the number of observers per galaxy with mass $M$ 
in each universe with $\neff$ and $\cc$. 
We define the density parameters as $\Omega_i = \rho_i /\rho_c$ ($i = {\rm rad}, \Lambda$),
where $\rho_c$ is the critical density of the present universe, and $\rho_i$ 
is evaluated when the energy density of dark matter in each universe 
becomes equal to the current density. 
The prior distribution $P_{\rm prior}$ depends on the production mechanism 
and will be discussed in the next section.

The number of observers $N_{\rm obs}$ in a galaxy is expected to be proportional to its mass $M$. 
We assume that $N_{\rm obs}$  is insensitive to $\eff$ and $\cc$, because
$N_{\rm obs}$ is determined locally in galaxies decoupled from cosmic expansion,
while $\eff$ and $\cc$ change only global properties of the universe.%
\footnote{
The change of $\eff$ affects the expansion rate at the BBN epoch and thus the primordial helium abundance. 
Since the stellar evolution depends on the initial helium abundance, the number of observers may depend on $\eff$. 
In the present analysis we drop the dependence assuming the change is minor. 
}
We also assume that 
the integral in \eq{Ptot1} is dominated by large galaxies 
with mass $M \gtrsim M_{\rm G} \sim 10^{12} M_\odot$ like the Milky Way. 
This is because the metals generated by the first-generation stars must be retained in the galaxy for the planetary formation. 
Under these assumptions, 
we can rewrite the probability as 
\beq
 P(\neff, \cc) \propto P_{\rm prior} (\neff, \cc) F (M > M_{\rm G}, \neff, \cc), 
 \label{Ptot}
\eeq
where $F$ is the fraction of matter that clusters into galaxies with mass larger than $M_{\rm G}$: 
\beq
 F (M > M_{\rm G}, \neff, \cc) \equiv \int_{M_{\rm G}}^\infty d M n_{\rm G} (M) M. 
\eeq
This can be estimated by using a spherical collapse model.

The observations of CMB 
revealed that primordial density perturbations are well approximated by a Gaussian. 
The time evolution of density perturbations can be studied by the linear perturbation theory. 
Hence it is reasonable to represent  
the distribution of density perturbations smoothed over a comoving scale $R$  by 
\beq
 P_\delta (R, t, \eff, \cc) \propto \exp \lkk - \frac{\delta^2}{2 \sigma^2 (R, t, \eff, \cc)} \rkk, 
 \label{Pdelta}
\eeq
where $\delta = \delta \rho/\rho$ is the matter density perturbation, 
and $\sigma$ is its variance. Note that the variance grows with time.

We are interested in the comoving scale $R_{\rm G}$ leading to the formation
of a galaxy with mass $M_{\rm G}$ where planets and observers are formed.
They are related by the mass conservation as 
\beq
 R_{\rm G} (M_{\rm G}) 
 &=& \lmk \frac{3 M_{\rm G}}{4 \pi \rho_{\rm m, 0}} \rmk^{1/3}
 \\
 &\simeq& 1.3 h^{-1} \Mpc \lmk \frac{\Omega_m h^2}{0.12} \rmk^{-1/3} 
 \lmk \frac{h}{0.7} \rmk \lmk \frac{M_{\rm G}}{10^{12} \Msolar} \rmk^{1/3}, 
 \label{R_G}
\eeq
where $\rho_{\rm m,0}$ and $\Omega_m$ are the present matter density and 
density parameter, respectively, and $h$ is the reduced Hubble constant. 
$M_G$ must be large enough to retain metals synthesized
in the first-generation stars for the subsequent formation of planets and life. 
It is not clear, however, which value of  $M_G$ is appropriate to use for the present analysis.
In the following we adopt $M_G = 10^{12} M_\odot$ as a reference value,
which is close to the Milky Way mass.
In some case we will also show the results for different values, $M_G = 10^6 M_\odot, 10^9M_\odot$, and 
$10^{13} M_\odot$, 
roughly corresponding to the masses of globular clusters, dwarf galaxies and galaxy groups, respectively.

\subsection{Evolution of density perturbations}

The variance of density perturbation smoothed over a scale $R$ is calculated from the power spectrum ${\cal P}_\delta (k)$ as 
\beq
 &&\sigma^2 (R, t, \eff, \cc) = \int_0^\infty \frac{4 \pi k^2 dk}{(2\pi)^3} {\cal P}_\delta (k) W^2 (k R), 
 \label{sigma-original}
 \\
 &&W (x) = \frac{\sin x - x \cos x}{x^3/3}, 
\eeq
where 
\beq
 &&\left< \delta (k) \delta^* (k') \right> = (2\pi)^3 {\cal P}_\delta (k) \delta^{(3)} (k - k'), 
 \\
 &&\delta (k) = \int d^3 x\, \delta (\vec{x}) e^{i \vec{k} \cdot \vec{x}}. 
\eeq
Note that the power spectrum ${\cal P}_\delta (k)$ is the Fourier transform of the correlation function for the density perturbation, which is different from $P_\delta (R)$ in \eq{Pdelta}.

From the Poisson equation, 
the density perturbation $\delta$ can be calculated from the gravitational potential $\Phi$ 
as 
\beq
 \delta (k, t) 
 &=& \frac{2}{3} \frac{k^2 a \Phi (k,t)}{\Omega_m H_0^2}, 
\eeq
The time-evolution and $k$-dependence of $\Phi$ are conveniently factorized as 
\beq
 \Phi (k, t) = \frac{9}{10} \Phi_p (k) T(\kappa) \frac{D(a)}{a}, 
\eeq
where $T(\kappa)$ is the transfer function, 
$D(a)$ is the growth function,%
\footnote{
We normalize $D$ such that 
$D = a$ during the matter dominated era, which is different from the one used in Refs.~\cite{Tegmark:2003ug, Pogosian:2004hd} 
by a factor of $2 a_{\rm eq} / 3$. 
We normalize the scale factor $a$ such that 
$a = 1$ at present when the matter energy density is equal to the observed value. 
}
and $\Phi_p$ is the primordial gravitational potential. 
The numerical factor $9/10$ represents the evolution of super-horizon modes 
around the matter-radiation equality.
The comoving wavenumber in the unit of a horizon scale at the matter-radiation equality, $\kappa$, is given by 
\begin{align}
 \kappa &= \frac{\sqrt{2}k}{a_{\rm eq}H (a_{\rm eq}) }
 =\frac{\sqrt{\Omega_{\rm rad}}}{\Omega_m} \frac{k}{H_0}, 
\end{align}
where $a_{\rm eq}$ ($= \Omega_{\rm rad} / \Omega_m$) is the scale factor at the matter-radiation equality. 
The matter power spectrum is then related to the  power spectrum of the primordial curvature perturbation ${\cal P}_\zeta$
as
\begin{align}
{\cal P}_\delta (k) = \frac{8\pi^2}{25} \frac{k}{\Omega_m^2 H_0^4} {\cal P}_\zeta(k) T(\kappa)^2 D(a)^2,
\end{align}
where
\begin{align}
{\cal P}_\zeta(k) &\simeq 2.101 \times 10^{-9} \left(\frac{k}{k_{\rm pivot}}\right)^{n_s-1}, 
\label{Pzeta}
\end{align}
with $k_{\rm pivot} = 0.05 {\rm Mpc}^{-1}$ and $n_s \simeq 0.965$~\cite{Aghanim:2018eyx}.

The transfer function describes the wavenumber dependence 
and the growth function describes the scale-factor dependence of the gravitational potential.
Here we briefly comment on the qualitative features of these functions. 
The density perturbation corresponding to the scale $R_{\rm G}$ enters the horizon 
before the matter-radiation equality. It is known that
the density perturbation at subhorizon scales grows only logarithmically 
during the radiation dominated era due to the Meszaros effect. 
The duration of this effect depends on the scale factor at the matter-radiation equality, $a_{\rm eq}$, 
and therefore $\delta \propto \ln \Omega_{\rm rad}$, 
where $\Omega_{\rm rad}$ is related to $\Delta N_{\rm eff}$ through \eq{Omega_rad}. 
On the other hand, the density perturbation grows as $a$ (i.e., $D(a) \propto a$) during the matter dominated epoch. 
For larger $\Omega_{\rm rad}$, the matter-radiation equality is delayed, and
the duration of the matter-dominated epoch decreases. Hence the density perturbation grows less until the present epoch. 
Combining these effects, we obtain $\delta \propto (\ln \Omega_{\rm rad}) / \Omega_{\rm rad}$. 
Below we will estimate $\delta$ (or $\sigma$) quantitatively 
and will see the result is consistent with this qualitative picture.

The fitting formula for the transfer function can be read from, e.g., Eq. (6.5.12) in Ref.~\cite{Weinberg:2008zzc}: 
\begin{align}
T(\kappa) & =  \frac{\ln \left(1+(0.124 \kappa )^2\right)}{(0.124 \kappa )^2}
\sqrt{\frac{1+(1.257 \kappa )^2+(0.4452 \kappa )^4+(0.2197 \kappa )^6}{1+(1.606 \kappa )^2+(0.8568 \kappa )^4+(0.3927 \kappa )^6}}. 
\label{T}
\end{align}
For the modes that enter the horizon 
before the matter-radiation equality, i.e., $\kappa \gg 1$, we obtain $T(\kappa) \propto \ln \kappa /\kappa^2$. 
The logarithmic dependence results from the Meszaros effect.

It is convenient to define a new time variable $x$ as 
\beq
 x \equiv \frac{\rho_\Lambda}{\rho_m (t)} = \frac{\Omega_\Lambda}{\Omega_m} ( 1+z)^{-3}.
\eeq
At the matter-radiation equality, 
it is given by 
\beq
 x_{\rm eq}^{-1/3} = \lmk x_{\rm eq}^{(\rm obs)} \rmk^{-1/3} 
 \lmk \frac{\Omega_\Lambda}{\Omega_\Lambda^{(\rm obs)}} \rmk^{-1/3} 
 \lmk \frac{\Omega_{\rm rad}}{\Omega_{\rm rad}^{(\rm std)}} \rmk^{-1}, 
\eeq
and $(x_{\rm eq}/x)^{-1/3} = (a_{\rm eq}/a)^{-1}$, 
where $( x_{\rm eq}^{(\rm obs)} )^{-1/3} \simeq 2820$ 
and $\Omega_\Lambda^{(\rm obs)} \simeq 0.69$~\cite{Aghanim:2018eyx}. 
The growth factor $D(a)$ is given by~\cite{Tegmark:2003ug}
\beq
 D(a) = \frac{2a_{\rm eq}}{3} \lkk 1 + \frac32 x_{\rm eq}^{-1/3} G(x) \rkk, 
 \label{D}
\eeq
where $G(x)$ is the growth factor in a flat universe filled with matter and vacuum energy, given by 
\beq
 G(x) &\equiv& \frac56 \lmk \frac{1+x}{x} \rmk^{1/2} 
 \int_0^x \frac{d x'}{x'^{1/6} ( 1+x')^{3/2}},
 \\
 &\approx& x^{1/3} \lmk 1 + \lmk \frac{x}{G^3 (\infty)} \rmk^\alpha \rmk^{-1/(3 \alpha)}.
 \label{G}
\eeq
Here the second line is a fitting formula with 
\beq
 &&\alpha = \frac{159}{200} = 0.795,
 \\ 
 &&G(\infty) = \frac{5 \Gamma(2/3) \Gamma(5/6) }{3 \sqrt{\pi}} \simeq 1.44. 
\eeq
For the scales of our interest, we can safely neglect the first term in \eq{D}.

The variance of the density perturbation after smoothing over a scale $R$ (see \eq{sigma-original}) is now given by 
\beq
 \sigma^2 (R, t, \eff, \cc) = \int_0^\infty d \ln k {\cal P}_\zeta (k) W^2 (k R) 
 \frac{4}{25} \frac{k^4 T^2(\kappa)}{\Omega_m^2 H_0^4} D^2 (a), 
 \label{sigma2}
\eeq
where ${\cal P}_\zeta (k)$, $T(\kappa)$, and $D(a)$ are given by 
Eqs.~(\ref{Pzeta}), (\ref{T}), and (\ref{D}), respectively. 
The dependence of the variance on the parameters can be read
by setting $k = 1/ R_{\rm G}$ in the integrand, and it reads
\beq
 &&\sigma (R, t, \eff, \cc)_{t \to \infty} 
 \nonumber\\
 &&\simeq 
 \sigma^{(\rm std)} (R_{\rm G}) 
 \lmk 1 + 0.18 \ln \frac{\Omega_{\rm rad}}{\Omega_{\rm rad}^{\rm (std)}} \rmk 
 \lmk \frac{\Omega_\Lambda}{\Omega_\Lambda^{(\rm obs)}} \rmk^{-1/3} 
 \lmk \frac{\Omega_{\rm rad}}{\Omega_{\rm rad}^{(\rm std)}} \rmk^{-1}
 \lmk \frac{G(\infty)}{G(x_p)} \rmk, 
 \label{simple}
\eeq
where 
\beq
 \sigma^{(\rm std)} (R_{\rm G}) \equiv \sigma (R_{\rm G}, t_p, \Delta N_{\rm eff}=0, \Omega_\Lambda^{\rm (obs)}) 
 \simeq 3.2, 
\eeq
is the variance at present ($t= t_p$) evaluated by the linear theory, \eq{sigma2}, 
and $x_p \equiv \Omega_\Lambda^{(\rm obs)} / \Omega_m$. 
The parameter dependence can be understood by noting how the duration of matter domination 
depends on the density parameters. That is to say,
the matter radiation equality is delayed if we increase the radiation energy. 
The cosmological constant comes to dominate earlier if we increase the cosmological constant. 
Since the matter density fluctuation grows efficiently only in the matter dominated epoch, 
the increase of the density parameters $\Omega_{\rm rad}$ and $\Omega_\Lambda$ suppress the growth of the density perturbations. 
The logarithmic dependence on $\Omega_{\rm rad}$ is due to the Meszaros effect.

\subsection{Anthropic bound}
When the density perturbation grows and exceeds the critical value $\delta_c$,
an overdense region collapses to form a galaxy.
The critical value can be calculated based on the spherical collapse model~\cite{Weinberg:1987dv} (see also Ref.~\cite{BarrowSaich}): 
\beq
 \delta_c \simeq \frac{9}{5} 2^{-2/3} G_\infty \simeq 1.63. 
\eeq
According to \cite{Martel:1997vi}, 
the fraction of matter that collapses into galaxies during the entire history of the Universe, $F$, 
is given by
\beq
 F (M > M_{\rm G}, \neff, \cc) 
 \propto 
 \int_\beta^\infty 
 d y \frac{e^{-y} }{s \sqrt{y} + \sqrt{\beta}},
 \label{F}
\eeq
where the parameter $\beta$ is given by 
\beq
 \beta \equiv \frac{\delta_c^2}{2 \sigma^2 (R_{\rm G}, t, \eff, \cc)_{t \to \infty} } \, . 
\eeq
Here, $s$ is a shape parameter 
that takes account of the fraction of the surrounding underdense region that also collapses into the galaxies. 
If we set $s \to \infty$, the result is proportional to the one given by the Press-Schechter formalism. 
We take $s = 1$, which is a reasonable case where the overdense region is surrounded by the underdense region with the same volume.

Assuming that the number of observers in a universe is proportional to the mass that collapses into galaxies, we can calculate the probability distribution of $\neff$ and $\cc$
 by using \eq{Ptot} and \eq{F}. 
The integral in \eq{F} is exponentially suppressed for $\beta \gg {\cal O}(1)$. 
This means that 
the fraction of matter that clusters into galaxies with $M > M_{\rm G}$ is exponentially 
suppressed for $\sigma \ll \delta_c$,
while it is of order unity for $\sigma \gtrsim \delta_c$. 
Roughly speaking, the condition $\sigma \gtrsim \delta_c$ is  the anthropic bound.
Since $\sigma$ depends on $\eff$ and $\Omega_\Lambda$, 
we can estimate their likely values 
that satisfy $\sigma \gtrsim \delta_c$. 
From the simplified expression \eq{simple}, 
we can see that $\Omega_\Lambda$ and $\Omega_{\rm rad} (\eff)$ 
cannot be much larger than the observed values 
from the anthropic argument.

\begin{figure}[t!]
\begin{center}
\includegraphics[width=95mm, bb=0 0 360 361]{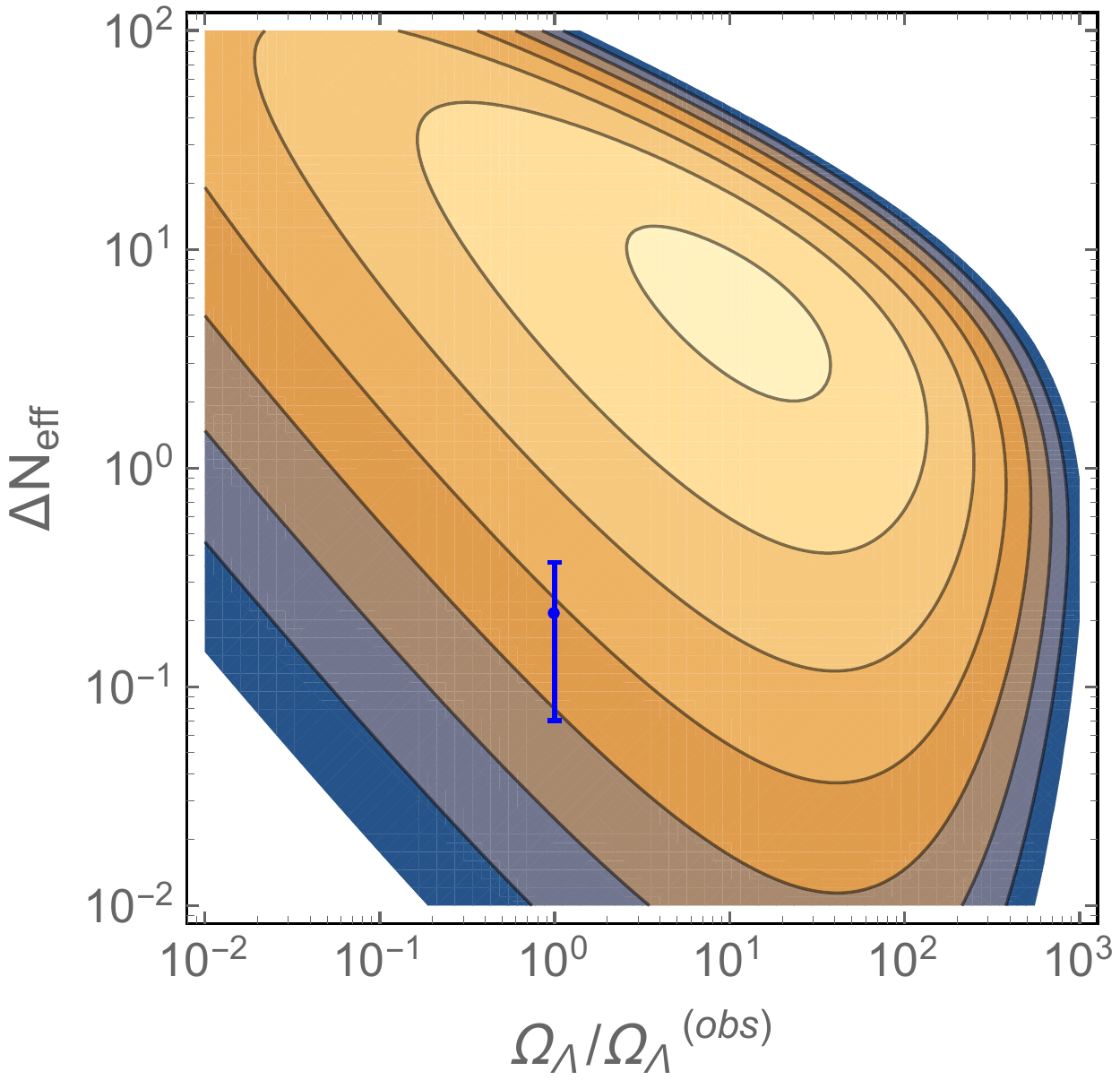}
\quad
\includegraphics[width=13mm, bb=0 -70 40 225]{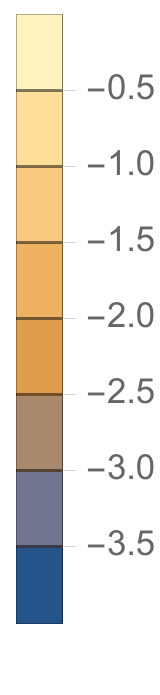}
\\
\vspace{1cm}
\includegraphics[width=95mm, bb=0 0 360 238]{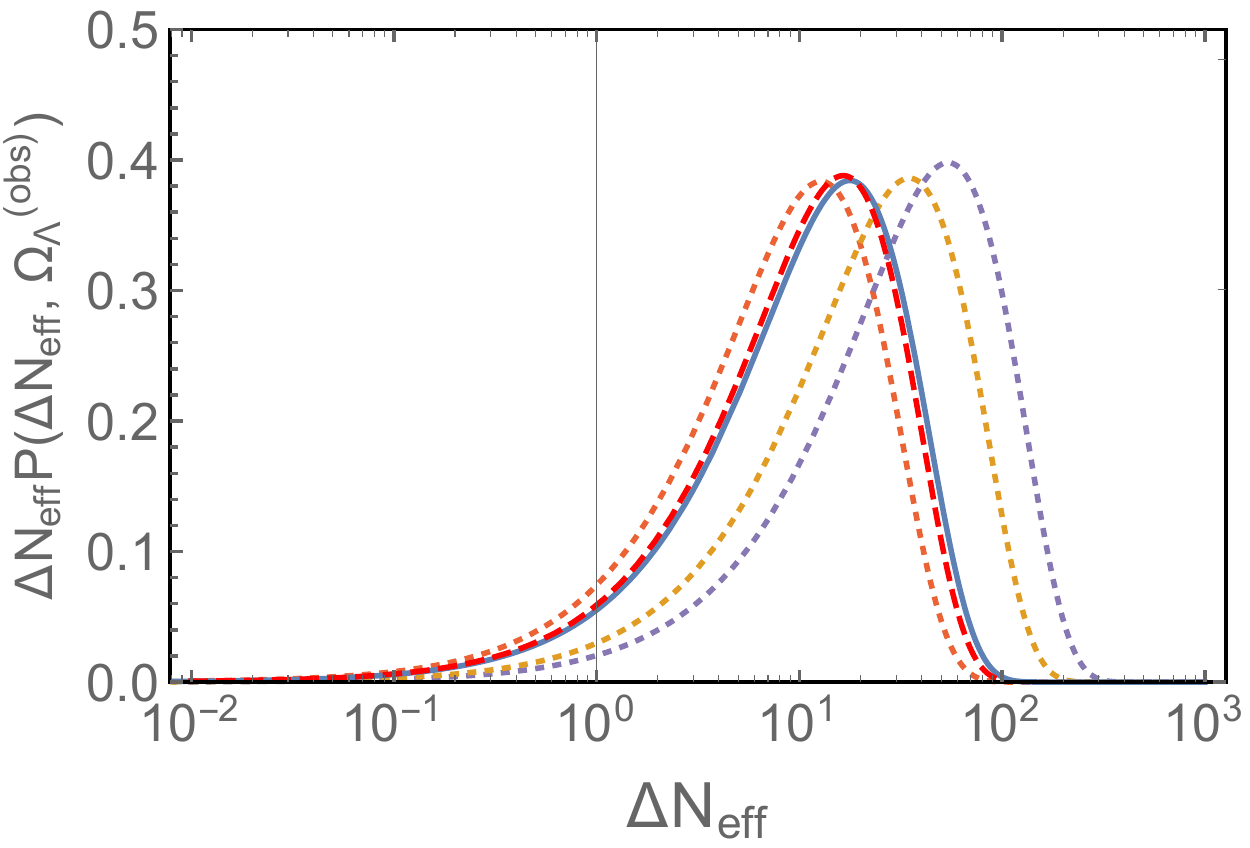}
\caption{
The probability distribution of parameters $\Delta N_{\rm eff}$ and $\Omega_\Lambda$ in the multiverse with 
a flat prior distribution, $P_{\rm prior} = 1$.
In the upper panel, we show a contour plot of $\log [ \Omega_\Lambda \Delta N_{\rm eff} P (\Delta N_{\rm eff}, \Omega_\Lambda)]$. 
In the lower panel, we show the normalized differential probability at $\Omega_\Lambda = \Omega_\Lambda^{(\rm obs)}$.
The blue solid line corresponds to $M_G = 10^{12} \M_\odot$.
The dotted lines correspond to 
$M_G = 10^{6} \M_\odot$, $10^{9} \M_\odot$, and $10^{13} \M_\odot$, respectively from right to left. 
The red dashed line is based on the simplified expression \eq{simple} with  $M_G = 10^{12} \M_\odot$, which is in good agreement with the blue solid one.
}
\label{fig1}
\end{center}
\end{figure}

The normalized probability distribution of $\Delta N_{\rm eff}$ and $\Omega_\Lambda$
is shown in Fig.~\ref{fig1}. 
Here we assume a flat prior distribution $P_{\rm prior}$ for both $\Delta N_{\rm eff}$ 
and $\Omega_\Lambda$ in \eq{Ptot} and set $M_G = 10^{12} M_\odot$ as a reference value.
In the upper panel, we show a contour plot of $\log [ \Omega_\Lambda \Delta N_{\rm eff} P (\Delta N_{\rm eff}, \Omega_\Lambda)]$. One can see that the most likely values of $\neff$ and $\cc$ are larger than those in our universe.
In the lower panel, we plot the probability distribution $\Delta N_{\rm eff} P (\Delta N_{\rm eff},  \Omega_\Lambda^{(\rm obs)})$ 
as a function of $\Delta N_{\rm eff}$, where 
the blue solid line is based on the numerical estimate of
\eq{sigma2}, while
the red dashed line is based on the analytic one \eq{simple}.
The two lines agree well with each other. One can also see that the typical value of $\Delta N_{\rm eff}$ 
is ${\cal O}(10)$.

The Planck data combined with the BAO observation 
gives the constraint~\cite{Aghanim:2018eyx} 
\begin{align}
N_{\rm eff} &= 3.27 \pm 0.15,
\end{align}
which is shown as the blue dot with an error bar in the upper panel of Fig~.\ref{fig1}. 
Interestingly, there is currently the so-called $H_0$ tension: the Hubble constant
inferred by the Planck and BAO (assuming $\Delta N_{\rm eff} = 0$) reads
$H_0 = (69.32 \pm 0.97) \ {\rm km/s/Mpc}$, while the local Hubble parameter measurement gives 
$H_0 = (74.03 \pm 1.42) \ {\rm km/s/Mpc}$~\cite{Riess:2019cxk}.
The significance of the tension is greater than $4 \sigma$.
In fact, $N_{\rm eff}$ and $H_0$ 
are correlated with each other in the Planck analysis; $\neff >0$ makes the
sound horizon smaller, which can be partially cancelled by larger $H_0$
because the last scattering surface becomes closer to us.
The tension can be relaxed if $\Delta N_{\rm eff} \gtrsim 0.4-0.5$.
The $H_0$ tension may hint at a sizable amount of dark radiation.

Now we shall discuss how likely 
the point $\Delta N_{\rm eff} = 0.5$ $(1)$ and $\Omega_\Lambda = \Omega_\Lambda^{(\rm obs)}$ 
are under the anthropic consideration. 
First, we note that 
the probability of finding ourselves in a universe with the present $\Omega_\Lambda^{(\rm obs)}$ or smaller
is about $3\%$ for the case of $\Delta N_{\rm eff} = 0$. 
We define 
the probability $\Delta N_{\rm eff} \le \Delta N_{\rm eff}^{(\rm max)}$ 
for $\Omega_\Lambda = \Omega_\Lambda^{(\rm obs)}$ 
by 
\beq
 N^{-1} 
 \int_0^{\Delta N_{\rm eff}^{(\rm max)}} d \Delta N_{\rm eff} 
 P (\Delta N_{\rm eff} , \Omega_\Lambda^{(\rm obs)}), 
\label{P1}
\eeq
where\footnote{
Precisely speaking, $\Delta N_{\rm eff}$ cannot be arbitrarily large as we assume a period 
of matter domination after the matter-radiation equality before the cosmological constant comes
to dominate the universe. This does not affect our results, though, because the number of
observers is significantly suppressed as the matter dominated epoch is shortened. 
} 
\beq
 N = 
 \int_0^\infty d \Delta N_{\rm eff} 
 P (\Delta N_{\rm eff}, \Omega_\Lambda^{(\rm obs)}). 
\eeq
Then we find that the probability to find ourselves in a universe with $\Delta N_{\rm eff} \le 0.5$ $(1)$
is about $0.03$ ($0.06$). See also the lower panel of Fig.~\ref{fig1}.
Thus we conclude that $\Delta N_{\rm eff} = 0.5$ (or $1$) is not unlikely based on the anthropic argument. 

When we vary both $\Delta N_{\rm eff}$ and $\Omega_\Lambda$, 
the probability to find ourselves in a universe with $\Delta N_{\rm eff} \le \Delta N_{\rm eff}^{(\rm max)}$ and $0< \Omega_\Lambda \le \Omega_\Lambda^{(\rm obs)}$ 
is given by 
\beq
 \int_0^{\Delta N_{\rm eff}^{(\rm max)}} d \Delta N_{\rm eff} 
 \int_0^{\Omega_\Lambda^{(\rm obs)}} d \Omega_\Lambda 
 P (\Delta N_{\rm eff} , \Omega_\Lambda). 
\label{P2}
\eeq
We find that this is about $0.003$ ($0.006$) for $\Delta N_{\rm eff}^{(\rm max)} = 0.5 (1)$ and $\cc^{(\rm obs)} = 0.69.$

The probability distributions for  $M_G = 10^6 M_\odot, 10^9 M_\odot$, and
$10^{13} M_\odot$ are also shown as dotted lines from right to left in the  lower panel of Fig.~\ref{fig1}. One can see that the probability
to find small values of $\Delta N_{\rm eff}$ increases as $M_G$ increases. Specifically, 
we find that the probability to find ourselves in a universe with $\Delta N_{\rm eff} \le 0.5$ $(1)$
is about $0.01$ ($0.02$) for $M_G =10^6 M_\odot$,
$0.015$ ($0.03$) for $M_G =10^9 M_\odot$ 
and $0.04$ ($0.07$) for $M_G =10^{13} M_\odot$.

The CMB-S4 experiment will improve the $1\sigma$ error for the dark radiation as $\delta (N_{\rm eff}) = 0.0156$~\cite{Wu:2014hta, Abazajian:2016yjj}. 
If the value of $\Delta N_{\rm eff}$ in our universe is determined by the anthropic principle, 
we would expect that the CMB-S4 experiment will find a nonzero value of $\Delta N_{\rm eff}$
close to the current upper bound.
On the other hand, if its result is consistent with $\Delta N_{\rm eff} =0$, 
we may conclude that the amount of dark radiation is not determined by the anthropic principle 
but is determined by some other mechanism. For example, 
the energy of inflation may dominantly converted to the SM particles at the time of reheating.

\begin{figure}[t!]
\begin{center}
\includegraphics[width=95mm, bb=0 0 360 364]{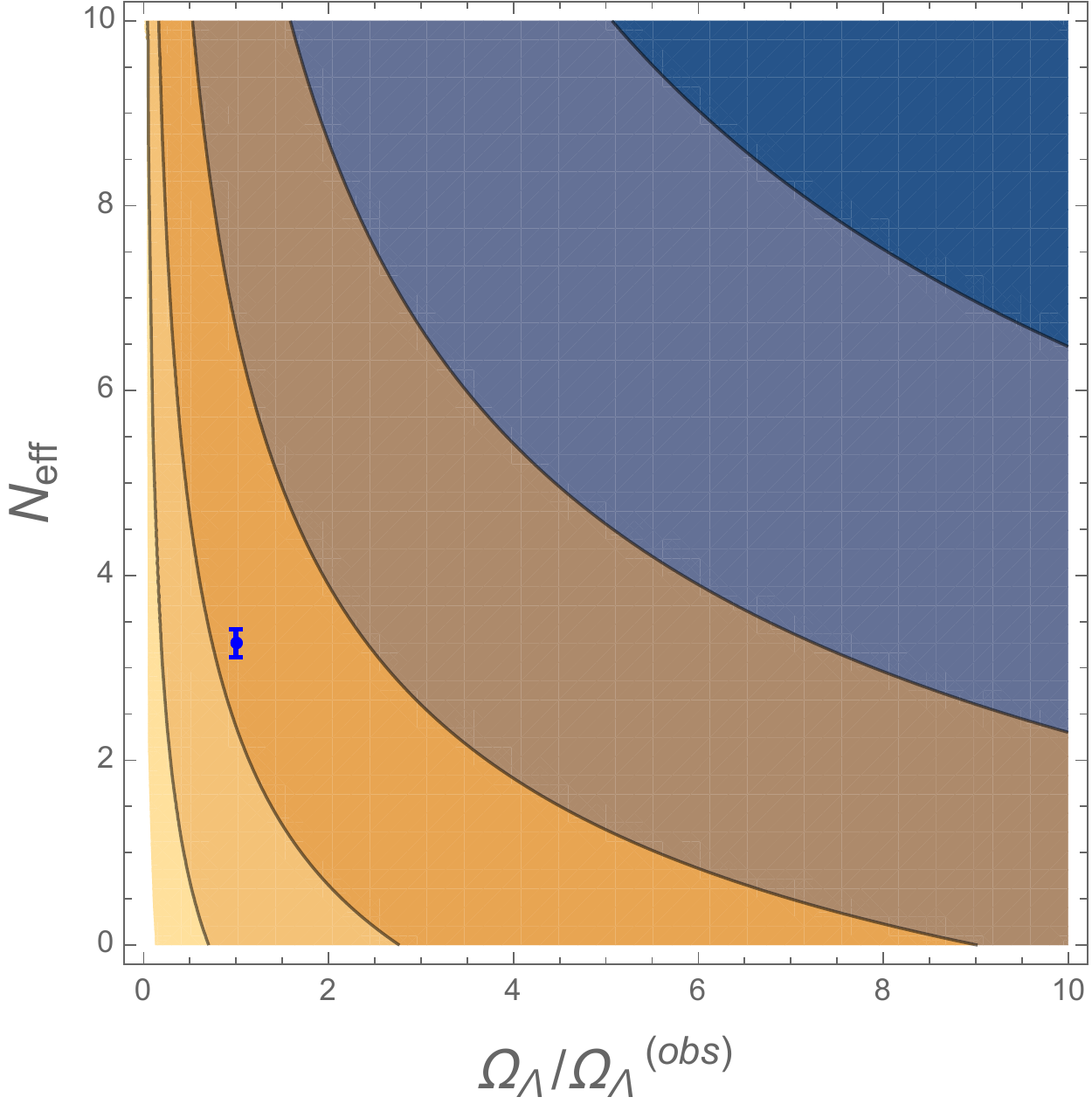}
\quad
\includegraphics[width=11mm, bb=0 -60 51 300]{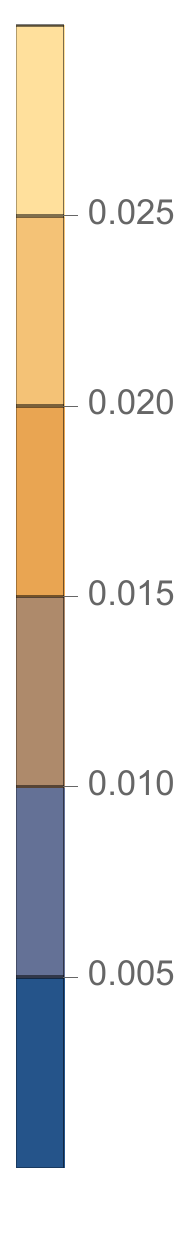}
\\
\vspace{1cm}
\includegraphics[width=95mm, bb=0 0 360 238]{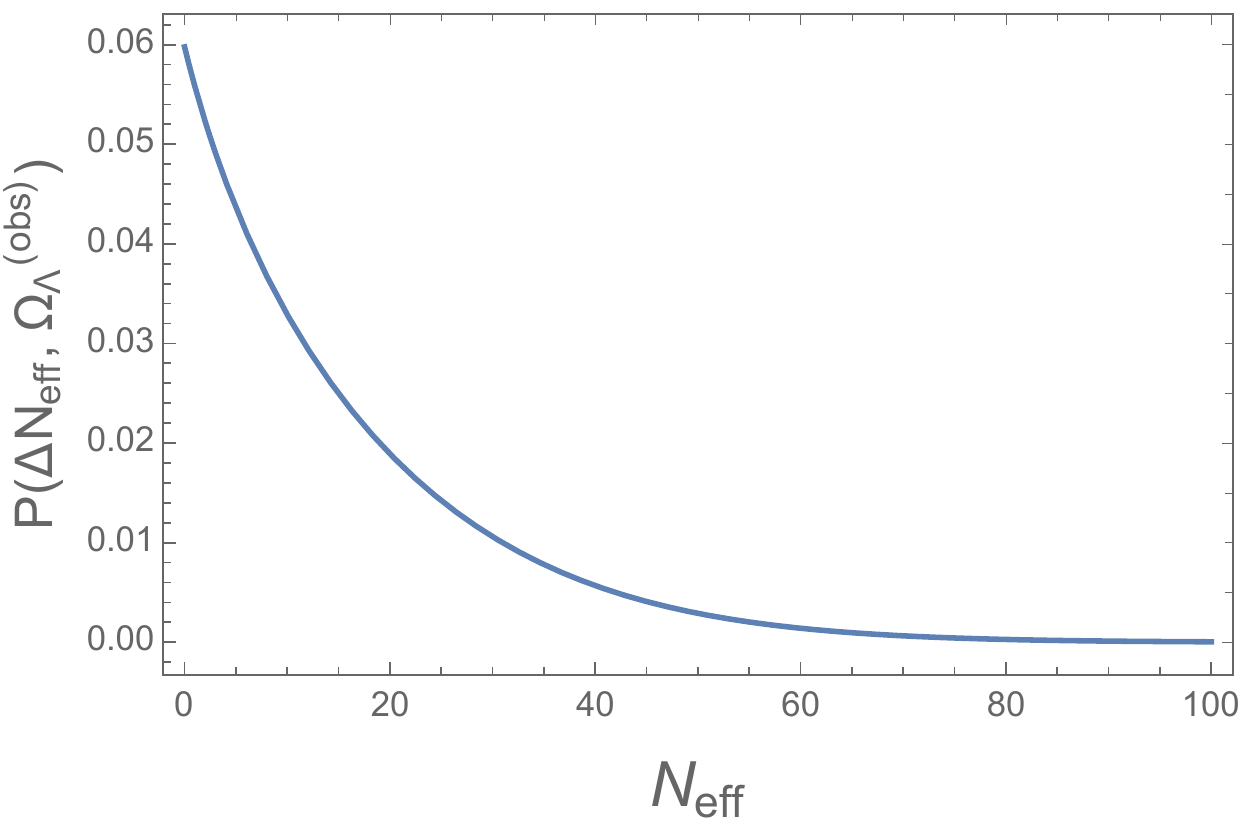}
\caption{
Same as Fig.~\ref{fig1} but with a linear plot for $N_{\rm eff}$ (instead of $\Delta N_{\rm eff}$) and $\Omega_\Lambda$. 
We set $M_G = 10^{12} M_\odot$. 
}
\label{fig-linear}
\end{center}
\end{figure}

Finally, we comment on the anthropic bound on the number of neutrino flavors $N_{\rm eff}$ instead of $\Delta N_{\rm eff}$.%
\footnote{
We thank Satoshi Shirai and an anonymous referee for raising this issue. 
}
The effective number of neutrinos $N_{\rm eff}$ can be smaller than the value in the standard cosmology, $N_{\rm eff}^{(\rm std)} \simeq 3.046$,  if the rehearing temperature is comparable to or lower than the neutrino decoupling temperature~\cite{Kawasaki:1999na,Kawasaki:2000en,Hannestad:2004px,Ichikawa:2005vw}. 
We can also consider a case in which a low-energy effective theory which is similar to the standard model but with a different number of generations is realized in the multiverse, and the number of generations may be considered as an environmental parameter. 
In the latter case, 
$N_{\rm eff}$ will be close to an integer number corresponding to the number of generations (if there is no dark radiation).
Motivated by such possibilities, we vary $N_{\rm eff}$ and $\Omega_\Lambda$  
assuming the flat prior distribution.
In Fig.~\ref{fig-linear} we show the probability distribution of $N_{\rm eff}$ and $\Omega_\Lambda$ in the linear plot . 
We find that the probability to find ourselves in a universe with $N_{\rm eff} \le 3$ 
is about $0.15$. 
Thus, the universe with three neutrino flavors is not unlikely at all based on the anthropic argument, if the prior distribution is flat.

%%%%%%%%%%%%%%%%%%%%%%%%%%%%%%%%%%%%%%%%%%%%%%%%%%%%%%%%%%%%%%%%
\section{Reheating and prior distribution
\label{sec:reheating}}
%%%%%%%%%%%%%%%%%%%%%%%%%%%%%%%%%%%%%%%%%%%%%%%%%%%%%%%%%%%%%%%%

In this section, we discuss a couple of simple models that predict dark radiation from reheating.
Suppose that the inflaton decays into dark radiation as well as the SM particles and 
that the dark radiation is completely decoupled from the SM sector. 
The extra neutrino species, which is proportional to the energy density of dark radiation, 
is then determined by the branching ratio into the dark radiation: 
\beq
 \Delta N_{\rm eff} = \frac{43}{7} \lmk \frac{43/4}{g_*} \rmk^{1/3} 
 \frac{\Gamma_D}{\Gamma_{\rm SM}}, 
\eeq
Here, we denote by $g_*$ ($\simeq 106.75$) the number of degrees of freedom of the SM particles at the time of reheating. 
The prior distribution of $\Delta N_{\rm eff}$ is then given by the 
probability distribution of $\Gamma_D/\Gamma_{\rm SM}$.

\subsection{Case of a single dark radiation component}

In superstring theories, scalar fields with flat potentials, called moduli, arise 
via compactifications on a Calabi-Yau space, and some of them may
be present in the low energy effective field theory~\cite{Candelas:1985en}. 
Inflation may be realized in the moduli space, and the decay of the inflaton
induces the reheating. 
Alternatively, coherent oscillations of moduli may dominate the energy density of the Universe 
after inflation and the subsequent moduli decay reheats the Universe. 
In either case the reheating occurs due to the moduli decay. 
In this section we focus on a single modulus that dominates the universe and decays into
the SM and dark radiation. 

The modulus $T$ has a shift symmetry along its imaginary component, the axion, which 
remains massless at the perturbative level.
We assume that the axion is almost massless, and so, once it is produced it contributes to dark radiation. 
This is the case if the real component of the modulus is stabilized by supersymmetry breaking effects.
Let us consider the following K\"ahler potential of the no-scale form: 
\beq
 K = - 3 \log \lkk T + T^\dagger 
 - \frac13 \lmk \abs{H_u}^2 + \abs{H_d}^2 + \lmk c_{\rm SM} H_u H_d + {\rm h.c.} \rmk \rmk \rkk 
 + \dots, 
\eeq
where we show only relevant terms and omit higher order terms responsible for e.g. the modulus stabilization,
and $c_{\rm SM}$ denotes a coupling constant. 
For simplicity, we assume that the superpotential and the gauge kinetic function 
are irrelevant for the modulus decay. 
Then the modulus decays only into the axion and the Higgs fields. The ratio of the decay rate 
is given by~\cite{Cicoli:2012aq,Higaki:2012ar,Higaki:2013lra} 
\beq
 \frac{\Gamma_D}{\Gamma_{\rm SM}} = \frac{1}{2 c_{\rm SM}^2}. 
\eeq
For a more generic K\"ahler potential, the modulus decay rate into axions is 
proportional to $(\del^3 K / \del T^3)^2$ ($\equiv c^2$), which may vary depending on the details of the
compactification etc. So let us parametrize the ratio as
\beq
 \frac{\Gamma_D}{\Gamma_{\rm SM}} = \frac{c^2}{c_{\rm SM}^2}, 
\eeq
where we take $c_{\rm SM} = {\cal O}(0.1)$.

\begin{figure}[t!]
\begin{center}
\includegraphics[width=105mm, bb=0 0 360 361]{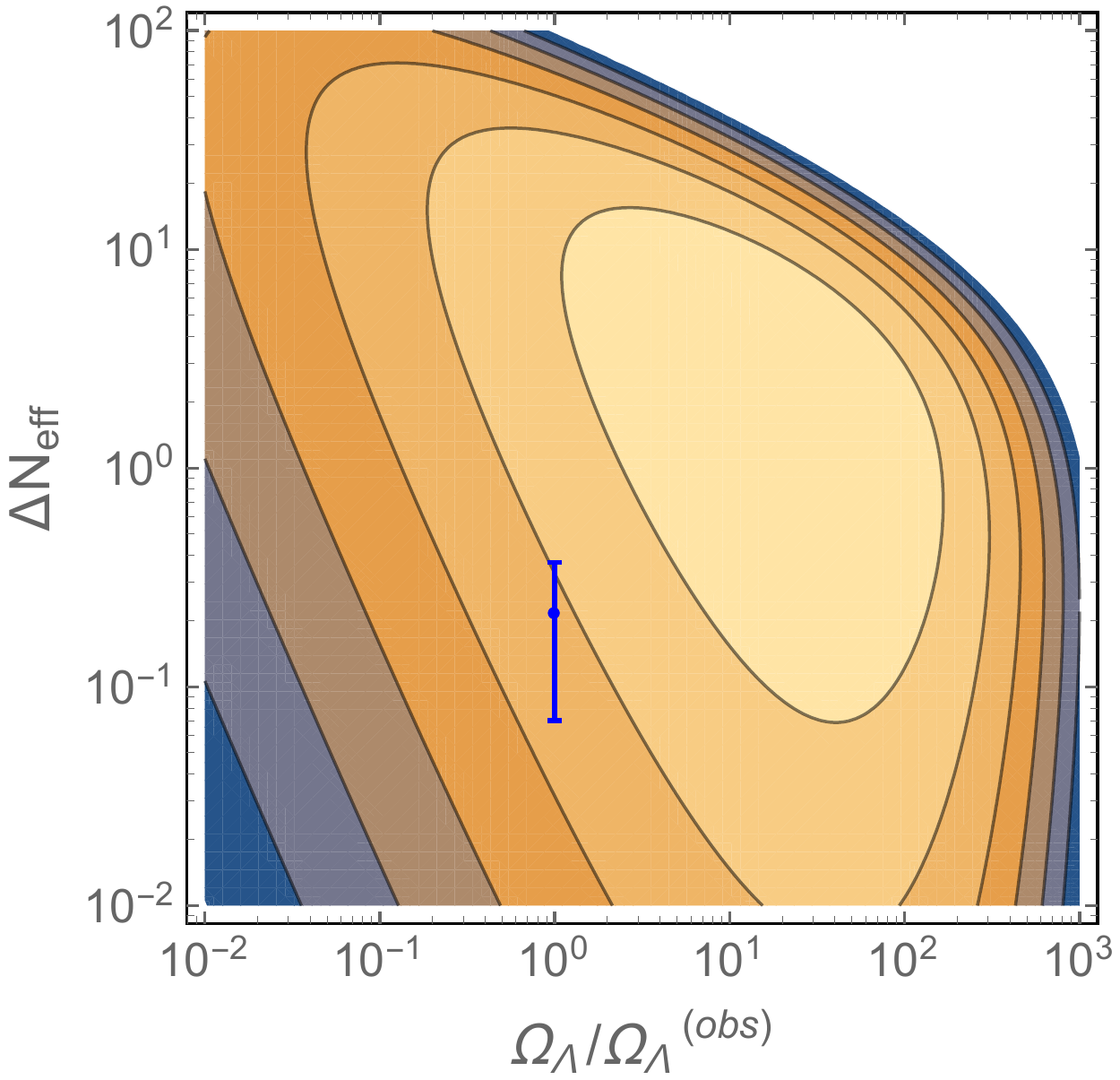}
\quad
\includegraphics[width=13mm, bb=0 -70 40 225]{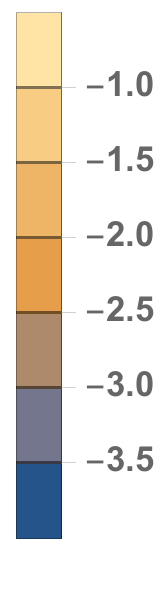}
\\
\vspace{1cm}
\includegraphics[width=105mm, bb=0 0 360 238]{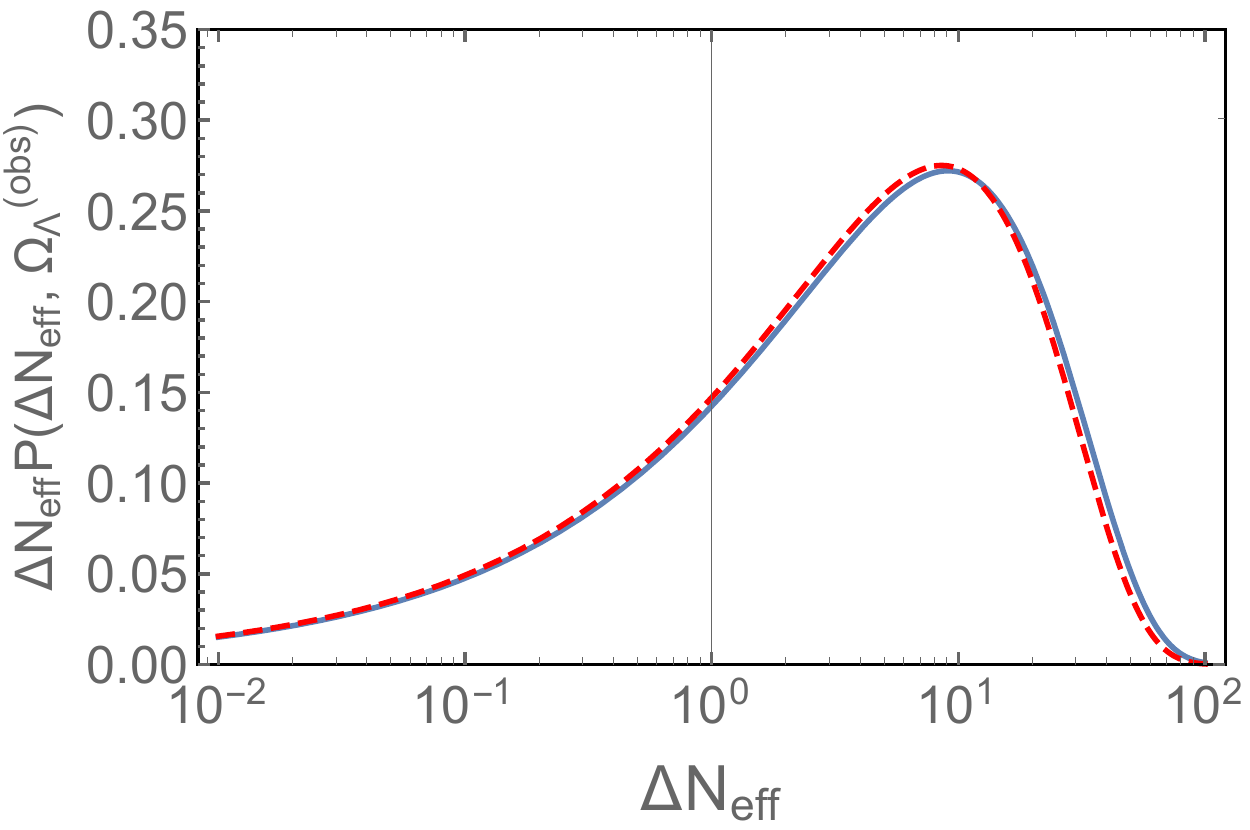}
\caption{Same as Fig.~\ref{fig1} but with $P_{\rm prior} \propto 1/\sqrt{\Delta N_{\rm eff}}$. 
}
\label{fig2}
\end{center}
\end{figure}

We assume that the coupling constant $c$ that determines $\Gamma_D$ is randomly distributed in the multiverse and its probability distribution is given by 
a flat distribution in the range of $\abs{c} \le \sigma$ ($= {\cal O}(1)$). 
We fix the decay rate into the SM particles for simplicity. 
Since the branching ratio into the dark sector is proportional to the coupling constant squared, 
the probability distribution of $\Gamma_D$ can be read from 
\beq
 P (c^2/\sigma^2) = 
 \left\{
 \bea{ll}
 \frac{1}{2 \sqrt{c^2/\sigma^2}} &~~~~{\rm for}~~ c^2 / \sigma^2 \le 1
 \\
 0 &~~~~{\rm for}~~ c^2 / \sigma^2 > 1,
 \eea
 \right.
\eeq
and is proportional to $1/\sqrt{\Gamma_D} \propto 1/\sqrt{\Delta N_{\rm eff}}$ for $c^2 / \sigma^2 \le 1$. 
Thus the distribution of $\Delta N_{\rm eff}$ is biased toward a smaller value. 
The probability distribution of $\Delta N_{\rm eff}$ and $\Omega_\Lambda$
is shown in Fig.~\ref{fig2} for the case of $P_{\rm prior} \propto 1/\sqrt{\Delta N_{\rm eff}}$.%
\footnote{
We implicitly assume that the typical value of $\Delta N_{\rm eff}$ is much larger than ${\cal O}(1)$ 
in the prior distribution. This is the case when $c / c_{\rm SM} \gg 1$. If this is not the case, 
the final distribution of $\Delta N_{\rm eff}$ is not strongly affected by the anthropic bound 
but is mainly determined by the prior distribution. 
}
We can see that the typical value of $\Delta N_{\rm eff}$ 
is ${\cal O}(1)$ in this case. 
The probability to obtain $\Delta N_{\rm eff} \le 0.5 (1)$ is given by 
$0.10 (0.14)$ based on \eq{P1}. If we also vary $\cc$, the probability
to obtain $\Delta N_{\rm eff} \le 0.5 (1)$ and $\cc \leq 0.69$ is $0.01 (0.02)$ based on \eq{P2}.

\subsection{Case of multiple dark radiation components}
We now consider how the probability distribution changes if there are multiple dark radiation 
components. In fact, the flux compactification of the higher-dimensional space in the string theory 
predicts a large number of axions and gauged dark sectors in the low-energy effective field theory. 
Inflation may occur in the axion field space, the so-called axion 
landscape~\cite{Higaki:2014pja, Higaki:2014mwa, Masoumi:2016eag, Masoumi:2017xbe, Yamada:2017uzq, Bachlechner:2018gew}.
For instance, the reheating could occur via the decay into gauge fields. 
If there are unbroken U(1) gauge fields in the dark sector, 
they contribute to dark radiation after the reheating. 
In this case, the number of particle species of the dark radiation, $N$, can
be larger than unity~\cite{Halverson:2019kna} and 
we parametrize the branching into the dark sector as 
\beq
 \frac{\Gamma_D}{\Gamma_{\rm SM}} = \frac{\sum_i c_i^2}{c_{\rm SM}^2}. 
\eeq

As in the previous case, we assume that the probability distributions of coupling constants $c_i$ are given by 
flat distributions in the ranges of $\abs{c_i} \le \sigma_i$ ($= {\cal O}(1)$). 
For simplicity, we set a universal value for the range, $\sigma_i = \sigma$. 
We also define $x \equiv \sum_i c_i^2 / \sigma^2 \propto \Gamma_D$.
Since $\Delta N_{\rm eff} \gtrsim {\cal O}(100)$ for $\sum c_i^2 \gtrsim 1$ and $c_{\rm SM} = {\cal O}(0.1)$,%
\footnote{
The explicit values of those parameters are not important for our discussion as long as the typical value of
 $\Delta N_{\rm eff}$ is larger than ${\cal O}(100)$, because of $P(\Delta N_{\rm eff}) \ll 1$
 for  $\Delta N_{\rm eff} \gtrsim {\cal O}(100)$. 
}
we are interested in the regime where $x \ll 1$. 
The probability distribution of $x$ is then calculated from%
\footnote{
The closed form of the probability distribution has been derived in Ref.~\cite{Rousseau} (see also Ref.~\cite{Forrester}). 
}
\beq
 P (x) 
 &=& \left. \frac{d}{d x} \int_{-\sigma}^\sigma \frac{d c_1}{2\sigma} \cdots \int_{-\sigma}^\sigma \frac{d c_N}{2\sigma} 
 \right\vert_{x < \sum_i c_i^2 / \sigma^2}
 \\
 &\simeq& 
 \frac{\pi^{N/2}}{2^N \Gamma (N/2)} x^{N/2-1}
 \qquad {\rm for} \ x \le 1. 
\eeq
Thus the prior distribution is almost flat for $N=2$, while it is strongly biases toward a large $\Delta N_{\rm eff}$ for $N>2$. 
In this case, the probability to obtain $\Delta N_{\rm eff} \le 0.5$ 
is strongly suppressed. 
Thus we conclude that the anthropic argument does not explain the current bound on $\neff$,
if the dark radiation that consists of $N (\gg 1)$ different particle species 
produced by the heavy scalar decay.

If one assumes that the probability distributions of the coupling constants $c_i$ are given by 
Gaussian distributions with zero mean and a universal variance $\sigma$, 
the probability distribution of $x \equiv \sum_i c_i^2 / \sigma^2$ ($\propto \Gamma_D$) is then given by the $\chi^2$-distribution with $N$ degrees of freedom: 
\beq
 P (x) = \chi^2 (N) = \frac{1}{2^{N/2} \Gamma (N/2)} x^{N/2 -1} e^{-x/2}. 
\eeq
Note that the result for the case of a single dark radiation component can be read from this formula by setting $N = 1$. 
For a small $x$, the probability distribution is proportional to $x^{N/2 -1}$. 
Since we are interested in a small $x$, 
the result is the same with that for the flat distribution.

%%%%%%%%%%%%%%%%%%%%%%%%%%%%%%%%%%%%%%%%%%%%%%%%%%%%%%%%%%%%%%%%
\section{Discussion and Conclusions
\label{sec:conclusions}}
%%%%%%%%%%%%%%%%%%%%%%%%%%%%%%%%%%%%%%%%%%%%%%%%%%%%%%%%%%%%%%%%

We have discussed the anthropic bound on the amount of dark radiation, assuming that 
the number of observers in each universe is proportional to 
the fraction of matter that clusters into galaxies with mass larger than the Milky Way galaxy. 
The matter-radiation equality is delayed if we increase the radiation energy. 
The matter density at subhorizon scales grows only logarithmically before the matter-radiation equality 
while it grows linearly in terms of the scale factor after that until the cosmological constant comes to dominate. 
As a result,  larger radiation energy leads to smaller density perturbations 
and hence a lower fraction of matter that clusters into galaxies. 
We have found that the number of observers is exponentially suppressed when 
the extra effective neutrino number exceeds of order $10$. 
If the prior distribution is flat, 
the probability to find ourselves in a universe with $\Delta N_{\rm eff} \le 0.5 (1)$ 
is about $0.03 (0.06)$, which is comparable to the probability to find  ourselves in a universe
with the observed cosmological constant or smaller. 
Therefore, the anthropic explanation of $\neff$ is not unlikely, if it is found to be around the
current upper bound.
We have also found that 
the probability to find ourselves in a universe with less than or equal to three 
neutrino flavors is about $0.15$ 
assuming the flat prior distribution in the multiverse.

We have also discussed a couple of examples in which dark radiation is produced
during the reheating process.
If a modulus is the inflaton or coherent oscillations of the modulus comes to dominate
the universe after inflation, the universe will be reheated by the modulus decay.
The modulus may also decay into dark radiation in addition to the SM particles.
For instance, if the modulus is stabilized by supersymmetry breaking effects, 
the modulus generically decays into its axionic partners with a sizable branching 
fraction~\cite{Cicoli:2012aq,Higaki:2012ar,Higaki:2013lra,Cicoli:2018cgu}. Alternatively,
the modulus may decay into multiple dark photons or axions. Assuming a flat prior
distributions for the coupling constants, 
we have found that the prior distribution of $\neff$ is proportional to $(\neff)^{N/2-1}$,
where $N$ is the number of particle species that constitute dark radiation.
In particular, if $N=1$, the energy density is biased toward smaller values 
and the probability to find ourselves in a universe with $\Delta N_{\rm eff} \le 0.5 (1)$ 
is about $0.10 (0.14)$. On the other hand, 
for $N \gg 1$, 
the prior distribution of $\neff$ is strongly biased toward larger values. 
In this case, 
the probability to find ourselves in a universe with $\Delta N_{\rm eff} \lesssim 1$ 
is strongly suppressed. In the latter case, some mechanism to dominantly reheat the 
SM sector may be required.

%---------------SECTION------------------%
%
\section*{Acknowledgments}
FT thanks Tufts Institute of Cosmology for warm hospitality, where
the present work was initiated. 
This work is supported by JSPS KAKENHI Grant Numbers
JP15H05889 (F.T.), JP15K21733 (F.T.),  JP17H02875 (F.T.), 
JP17H02878(F.T.),  and by World Premier International Research Center Initiative (WPI Initiative), MEXT, Japan.
%
%---------------SECTION------------------%

\vspace{1cm}

\bibliography{reference}

\end{document}